# Resilience Enhancement at Edge Cloud Systems

**Jose Moura[1] and David Hutchison[2]**
[1]ISCTE-Instituto Universitário Lisboa, Instituto de Telecomunicações, Lisbon, Portugal
[2]School of Computing and Communications, Lancaster University, Lancaster LA1 4WA, U.K.

Corresponding author: Jose Moura (e-mail: jose.moura@iscte-iul.pt)

The work of Jose Moura was supported by Fundação para a Ciência e Tecnologia / Ministério da Ciência, Tecnologia e Ensino Superior (FCT/MCTES) through National Funds When Applicable Co-Funded European Union (EU) Funds under Project UIDB/50008/2020; and in part by the Instituto de Telecomunicações, Lisbon, Portugal. The work of David Hutchison was supported in part by the Next Generation-Converged Digital Infrastructure (NG-CDI) Project through U.K. Engineering and Physical Sciences Research Council (EPSRC) under Grant EP/R004935/1, and in part by British Telecom plc.

**ABSTRACT** It is becoming common practice to push interactive and location-based services from remote datacenters to resource-constrained edge domains. This trend creates new management challenges at the network edge, not least to ensure resilience. These challenges now need to be investigated and overcome. In this paper, we explore the use of open-source programmable asset orchestration at edge cloud systems to guarantee operational resilience and a satisfactory performance level despite system incidents such as faults, congestion, or cyber-attacks. We discuss the design and deployment of a new cross-level configurable solution, Resilient Edge Cloud Systems (RECS). Results from appropriate tests made on RECS highlight the positive effects of deploying novel service and resource management algorithms at both data and control planes of the programmable edge system to mitigate against disruptive events such as control channel issues, service overload, or link congestion. RECS offers the following benefits: i) the switch automatically selects the standalone operation mode after its disconnection from the upper-level controllers; ii) deployment of edge virtualized services is made, according to client requests; iii) the client requests are served by edge services and the related traffic is balanced among the alternative on-demand routing paths to the edge location where each service is available for its clients; iv) the TCP traffic quality is protected from unfair competitiveness of UDP flows; and v) a set of redundant controllers is orchestrated by a top-level multi-thread cluster manager, using a novel management protocol with low overhead.

## I. INTRODUCTION

Programmable networking concepts including Software Defined Networking (SDN) offer excellent prospects for highly adaptable and fast management of network resources and data flows [1] in many modern networking scenarios such as Internet of Things (IoT)-enabled healthcare systems [2] or Peer-to-Peer (P2P) energy trading in intelligent transportation systems [3]; these are typically hybrid systems joining communication and computational edge resources. Nevertheless, SDN also brings potential problems such as increasing management complexity and as a target for attack, potentially compromising the desired resilience of such systems [4][5]. In fact, an SDN-based system could experience significant degradation of its performance due to various system threats or congestion at both data forwarding and control levels. Consequently, it is appropriate to explore suitable programmable solutions to mitigate unexpected system faults [6], congestion [7], or cyber-attacks [8].

### A. BACKGROUND

The background of current research work as well as the main operating scenario are now presented. The emerging data-intensive, interactive, and location-sensitive user services such as 5G, IoT, augmented reality, and vehicle-to-vehicle communications [9] force computational resources being moved from remote clouds [10] to edge clouds [11] in order to diminish the data and service access latency, to provide the edge network infrastructure with local scalable processing, and even to run local self-adaptable algorithms [12]. The higher heterogeneity and scarcity of edge resources must not impair the upcoming demand for reliable and efficient edge services. Therefore, we propose RECS (Resilient Edge Cloud Systems), a programmable serverless system [13] that orchestrates both elastic networking and computing resources for enhancing the operational resilience of local on-demand virtualized services at edge network domains, shown in Fig. 1.

## B. CONTRIBUTION AND NOVELTY

This work proposes practical solutions to mitigate the negative effects on the system performance imposed by system menaces or high loads. These solutions are deployed with a cross-layer design at distinct vertical levels of the SDN-based system (see Fig. 2), viz. data forwarding, switching logic, and resilience management, aiming to build up a more resilient SDN-assisted system for edge computing network scenarios. In a nutshell, the main aims of this contribution are to design, implement, and evaluate diverse programmable solutions to satisfy a set of goals, as follows: i) increase the resilience of a SDN-based system after communication failures between network devices and their controllers by using a suitable switch functional mode; ii) mitigate congestion situations at the network server side, providing an elastic supply of virtualized services that follows load variation; iii) overcome congested links and collaborate with other solutions towards the most effective use of system available resources; and iv) balance control workload among SDN controllers and increase resilience at the system control level. By coherently integrating diverse programmable solutions, our SDN-based work is a novel orchestrator for edge cloud distributed resources, enhancing system operational resilience in challenging cases such as faults, congestion, or attacks. We also envisage the minimization of deployment and operational costs of the network infrastructure owned by a specific Internet provider.

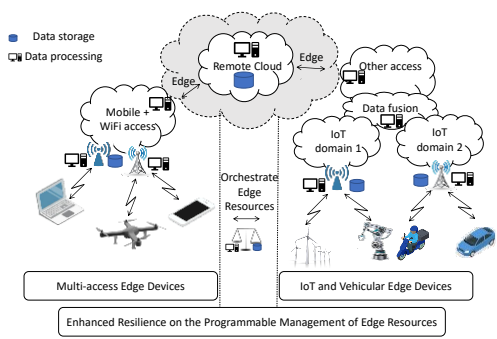

**FIGURE 1.** RECS - a programmable serverless computing system providing on-demand edge resources and elastic virtualized services with enhanced resilience

## C. STRUCTURE

The paper structure is as follows. After the introduction, Section II analyses related work, highlighting the novel aspects of our work. Section III discusses the design of RECS. The deployment of RECS is in Section IV, and Section V broadly evaluates the proposed system. Section VI summarizes the discussion about the main evaluation results and describes how our proposal can be applied to real-life scenarios. Finally, Section VII concludes the paper with some suggestions for future research directions.

## II. RELATED WORK

Recent advances have occurred in the resilient operation of programmable edge systems [14][15][11][16]. The work reported here builds on the authors' previous publications [14][15] which investigate important design steps towards resilient operation of programmable systems at the network edge, in the face of severe threats to their normal operation. Specifically, [14] discusses cooperation models among system players together with a penalization mechanism against defecting players, aiming for resilient operation. The main outcome from [15] is that the offloading of processes and data in (edge) cloud-based scenarios may undermine the accuracy of anomaly system detection components unless proper corrective actions are taken to increase the robustness of these components. Ref. [11] discusses best practices, following European Telecommunications Standards Institute (ETSI) standards, to undermine the negative impact induced by core and access threats on the performance of Multi-Access Edge Computing systems as well as to preserve the privacy of mobile users. The work in [16] thoroughly investigates the interaction between SDN and edge computing. Using this technological synergy some key benefits are obtained such as bringing low-cost computational solutions into the proximity of edge devices. Nevertheless, there are open issues, viz: management complexity, mobility, energy and computational constrained edge assets, heterogeneity, scalability, reliability, and security. In our work, we aim to study the open issues associated with system reliability and system scalability.

We have reviewed the literature for SDN-based techniques to mitigate the negative performance effects induced by system faults and congestion, i.e. two main perspectives of our current work. On one hand, our literature analysis identifies recent work [17][18] that deals with fault management. Ref. [17] uses a synchronized mechanism to periodically update the controller's state among a set of SDN controllers. In case of failure of the current responsible controller, the same mechanism can select another working controller based on the distance and delays among different network entities. The authors of [18] propose a SDN based fault-tolerant routing architecture for IoT environments. Their solution discovers redundant and non-overlapping routing paths between network equipment by using link costs. The cost of each link considers both the percentage of link usage and the rate of link delay.

On the other hand, we have assumed load balancing techniques [19][20][21][22][23] as a possible way to mitigate the negative performance effects imposed by congestion in programmable networks. The authors of [19] propose a deep reinforcement learning-based routing scheme aimed at balancing the load among the network links. In addition, [20] investigates a SDN-based solution to balance the service load amongst data plane servers. Ref. [21] scales out the control channel load across the diverse SDN controllers.

The authors of [22] propose an algorithm that enables the SDN controller to select, from a server farm, the server with the most suitable service response time to satisfy a specific client request. Nevertheless, this proposal suffers from system bottleneck and the issue of a single point of failure because it uses only a single SDN controller. As an alternative, we propose a solution with multiple controllers to enforce both control robustness against failures and control scalability. In addition, we propose a cluster manager at the resilience management level that orchestrates the underlying operation of the diverse SDN controllers. This orchestration is made via a novel signaling protocol detailed in sub-sections III.D and IV.D, and tested in sub-section V.D. The literature also points out many load balancing metrics as follows [23]: throughput, peak load ratio, utilization, response time, overhead, root mean squared error, packet loss rate, percentage of matched deadline flow, energy consumption, migration cost, execution time, load balancing degree, guaranteed bit rate, overload ratio, average number of synchronizations, workload, cumulative frequency, and latency. Our work uses the workload metric to coordinate SDN controllers, and the load detection metric to balance traffic at the data forwarding level.

We discuss below previous work on the control of programmable systems, concerning fault tolerance [24] and the orchestrated control [25][26] among a set of SDN controllers. The authors of [24] propose a master-slave protocol that replicates the control logic from a single controller to other backup controllers for fault tolerance, without requiring any code changes to the initial controller. The proposal adds explicit acknowledgement messages to the OpenFlow protocol and deploys buffers on existing switches for event retransmission and command filtering. These changes are necessary to support the exactly-once control action at the switches after any system failure sequence. Nevertheless, the need for previous alterations can be a serious handicap to using that solution in real deployments. Alternatively, our proposal (debated in sub-sections III.D and IV.D) orchestrates the control logic among several SDN controllers without modifying OpenFlow and switch code.

The work in [25] proposes a workload balancing mechanism for distributed SDN controllers, where idle controllers assume the control workload of overloaded controllers. The authors in [26] go a step further by suggesting a more flexible and elastic distributed controller design in which the number of running controllers within a cluster follows the data plane traffic load. Our proposal also offers an elastic behavior not at the control level as in [26], but in our case at the data forwarding level, activating servers from a server farm as needed. This is completely aligned with the model of serverless edge computing [13].

Table I summarizes related discussed work and highlights the novelty of our RECS research that, to our knowledge, is the first to propose consolidation in a single solution all the following aspects: programmable edge system, fault management, orchestration of redundant controllers, load balancing, and on-demand activation of virtual services.

TABLE I. LITERATURE COMPARISON (+ ASPECTS COVERED)

| Ref. | SDN | Novel issues induced by SDN and edge computing | Fault management | Orchestration | Load balancing | On-demand activation |
|---|---|---|---|---|---|---|
| [13] | | | + | + | | + |
| [14] | + | | | + | | |
| [15] | + | | + | | | |
| [11][16] | + | + | | | | |
| [17][18] | + | | + | | | |
| [19][20][21][22][23] | + | | | | + | |
| [24] | + | | + | | | |
| [25] | + | | | + | | |
| [26] | + | | | + | | + |
| RECS | + | + | + | + | + | + |

## III. DESIGN

Now we present several cross-level design solutions for increasing the resilience of an SDN-based system in adverse situations such as system faults, congestion, or attacks. These solutions are aggregated in our RECS (Resilient Edge Cloud Systems) proposal. RECS has four parts: i) the first describes the design of a solution whereby switching devices autonomously detect and react against communication failures to the control level; ii) the second part shows how a SDN controller can behave as a Proxy-ARP for ARP requests. This Proxy-ARP operation can balance a high number of service requests from clients among the available topological servers of an elastic edge server farm; iii) the third part presents how to architect a flexible and programmable solution to overcome congested links using Select groups at the data message forwarding level; and iv) the last part studies the workload orchestration among SDN controllers and the associated system overload. The RECS overall design layout is visualized in Fig. 2. This design aims to enhance the system's operational resilience against both communications failures and congestion in edge computing domains. It has three functional levels. The data forwarding (lowest) level contains virtualized servers, service clients, and switching devices controlled by (higher level) SDN controllers. These redundant controllers form the intermediate and reliable switching logic level. The cluster manager of the resilience management (topmost) level orchestrates the SDN diverse controllers of the intermediate functional level.

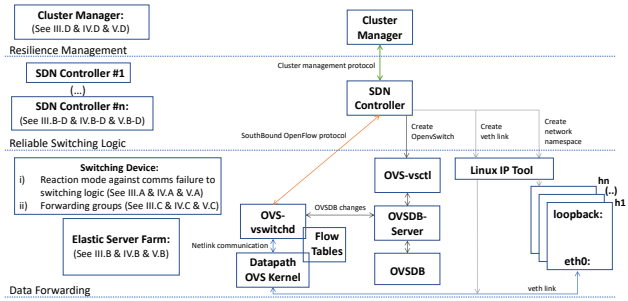

FIGURE 2. Architectural layout of the proposed cross-level programmable solution (RECS) to enhance the system's operational resilience against both communications failures and congestion in edge computing domains. Inside each block at the left are the paper sub-sections that discuss the key block functionalities and their evaluation

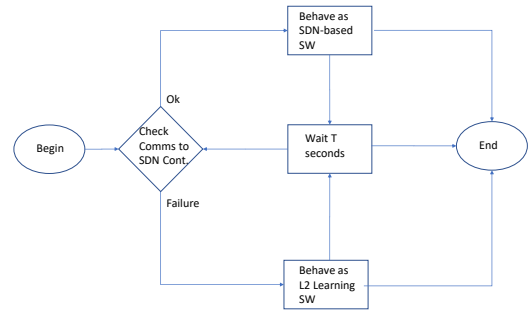

FIGURE 3.b OpenvSwitch standalone operation mode after the switch becomes disconnected from the upper SDN controller

## A. ENHANCING SYSTEM RESILIENCE AGAINST CONTROL CHANNEL COMMUNICATION FAILURE

We now discuss the most appropriate functional mode of an SDN-based switch after a communication failure between that switch and the SDN controller. Considering a software-based switch, such as OpenvSwitch, there are two alternative functional modes: secure (Fig. 3.a.) or standalone (Fig. 3.b.). When the switch is on the secure mode, after a communication failure with the controller, the switch discards any received message through any input port. In addition, the switch tries to reconnect with the SDN controller. Alternatively, when the switch is on the standalone mode and lacking communication with the controller, the switch behaves like a L2 MAC learning autonomous switch. Also, in the standalone mode, the switch tries to re-establish communication with the SDN controllers. Considering the two discussed functional modes and that we are mainly interested on a resilient operation of the SDN-based system, the standalone is the preferable mode, because switches can operate independently of controllers.

## B. AVOIDING SERVER CONGESTION

To avoid the server congestion problem, we virtualize elastically a cluster of servers at the data forwarding level and, with the help of the SDN controller we divert the traffic from each client destined to a specific service towards a server different from the alternative servers available to other clients. Thus, we balance the load of all clients among the diverse servers available in the cluster (i.e. the server farm). The design of this solution is summarized in Fig. 4.

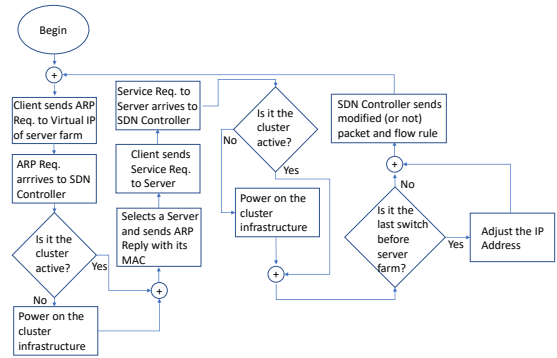

FIGURE 4. Flowchart of SDN-based system involving clients, switches, SDN controllers, and elastic servers particularly focused in the communication between the clients and the server farm

The current solution manages both the ARP protocol and the service request message, making some IP address changes in the message header like a home-based wireless NAT router. The first functional step of our solution is triggered by the arrival of an ARP Request to a switch. Considering the switch cannot make a positive match with any local flow rule, then the switch, following a default flow table-miss rule, it sends a copy of the received message to SDN controllers. Then, each SDN controller verifies first if it should decide about how to control that ARP Request (see sub-section III.D about the coordination among SDN controllers). When the copy of the received ARP Request is processed by an elected SDN controller, that controller chooses the server (i.e. the physical MAC address of the server network interface) from the pool of servers which are offering the same service to the potential clients. The controller should verify if the selected server is already in operation. Otherwise, the SDN controller should activate that server before sending to the client a new created ARP Reply message associating the IP_VIRTUAL of the required server farm service with the MAC of the selected server. In this way, the SDN controller acts as a Proxy-ARP.

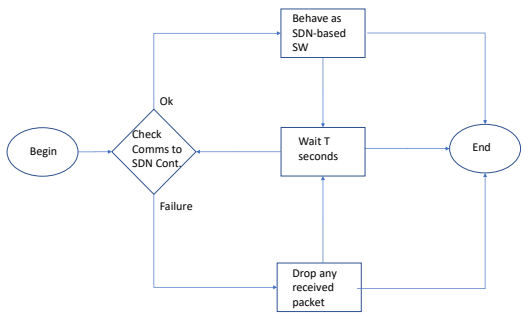

FIGURE 3.a OpenvSwitch secure operation mode after the switch becomes disconnected from the upper SDN controller

The second functional step (see Fig. 4) is when the SDN controller receives a packet with a destination IP address equal to IP_VIRTUAL. At this moment, the SDN controller verifies again if the cluster infrastructure is in operation. This new verification in the SDN controller is important to support a correct functional system behavior because after the last infrastructure activation the same server farm infrastructure, due to an idle timer, could have been switched off and, at the same time, the client ARP table still locally holds the last selected MAC address of the server farm. Then, the SDN controller should be ready to activate the server farm infrastructure in the presence of either an ARP Request or another IP packet. After the previous verification was made, the SDN controller verifies if the switch that has produced a control processing event is the last switch at the communication path to the server farm. If this is true, the SDN controller (acting as a NAT router) replaces the server farm virtual IP address by the currently used IP address of the selected server. In the case of the reply message to the previous packet (when it is the first switch from the topological perspective of the server farm), the SDN controller modifies the source IP address of the answering packet. This bidirectional modification on IP addresses made by the SDN controller is like a middleman attack on an authentication security protocol, but in the current scenario for a "good" cause. In this way, from the client perspective, the client gets the feeling he/she is interacting with a server via the IP address IP_VIRTUAL and, at the other end, the server perceives the client has tried to contact that server using (as normally expected) the IP address assigned to the network interface of that server. The main advantage of this IP address modifying solution is to enable the SDN controller, using a programmable criterion, to balance the load of many clients' requests to a single service among any number of elastic servers of a server farm offering that service. In addition, the processing of the SDN controller described in this sub-section can be also successfully employed in a distinct scenario from the current one – namely service access control or protection against cyberattacks. Then the controller, before replacing the IP address of each initial flow packet, can verify whether the client has enough privileges to use the requested service or if the packet belongs to a legitimate flow. In either case, when the SDN controller concludes that the client is not allowed to use the service or the received packet belongs to an ongoing attack, the controller can drop the current received packet and even install flow rules in the switches to drop the subsequent packets of flows that cannot be forwarded.

## C. OVERCOME LINK CONGESTION AND PROTECT TCP QUALITY AGAINST UDP RESOURCE USAGE UNFAIRNESS

We use here the OpenFlow Select group configured with the help of the open-source NetworkX Python library to reduce link congestion. The OpenFlow Select group is primarily designed for load balancing at the switch via multipath routing to the same destination. In addition, the usage of this group mitigates the negative impact of non-stopping data forwarding traffic loops. As indicated in Fig. 5, each bucket in a Select group has an assigned weight, and each packet that enters the group is sent to a single bucket. There are several possible ways to select the more feasible bucket for every message flow. Each switch's implementation imposes the bucket selection method to be used. For example, in the case of OpenvSwitch, the bucket of a Select group can be selected as follows. In OpenvSwitch 2.3 and earlier, OpenvSwitch used the destination Ethernet address to choose a bucket in a select group. In a different way, OpenvSwitch 2.4 and later by default hashes the source and destination Ethernet address, VLAN ID, Ethernet type, IPv4/v6 source and destination addresses, and protocol. Specifically, for TCP segments, the source and destination ports can be also used to select the bucket.

The bucket weights allow the selection of a specific bucket among others. Each bucket in a Select group has a list of actions. These actions are supported by OpenFlow. In our design, the bucket weights are evaluated using the lowest cost path (i.e. Dijkstra's shortest path algorithm from NetworkX) to each possible destination. The costs of links are dynamic because these are evaluated by the SDN controller using for example transmission link rates. These rates are evaluated from OpenFlow statistics periodically retrieved by the controller from all the switches, as shown in Fig. 6. In phase 1, the controller collects port and flow rule statistics from the data plane switches.

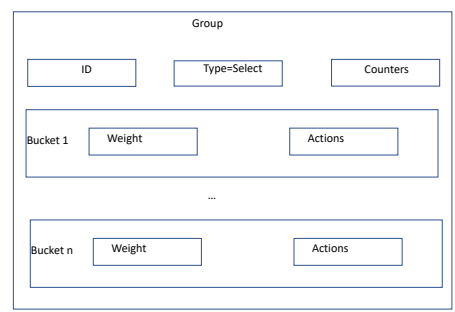

**FIGURE 5.** Each Select group is formed by several buckets. Each bucket has associated a list of OpenFlow actions

Then, the controller uses the received statistics of phase 1 and updates (in phases 2 and 3 respectively) the link costs and bucket weights. Finally, in step 4, the controller transfers Select groups to the switches with new bucket weights reflecting the last status reported from the data plane. This solution to overcome link congestion also protects the TCP traffic quality against the unfair network resource usage by UDP competitive traffic (see sub-sections IV.C and V.C).

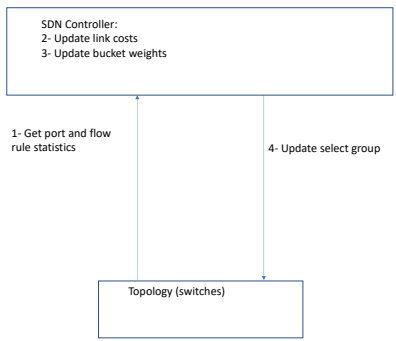

**FIGURE 6.** Design of solution to mitigate link congestion based on the collection of switch statistics by the SDN controller, update of link costs and bucket weights, and transfer of renewed select groups from the SDN controller to the controlled switches

### D. ORCHESTRATING A CLUSTER OF SDN CONTROLLERS AND THE CONTROL CHANNEL LOAD

We discuss the diverse roles each SDN controller can assume in SDN configurations with multiple controllers, assuming the SouthBound (S/B) API is using the OpenFlow protocol. In this type of scenario, each SDN controller could assume one of three possible roles in relation to each switch on the data forwarding system level: master, slave or equal. A SDN controller with the master role monitors and controls the data forwarding switching devices; it can also process asynchronous S/B messages such as Packet In. On the other hand, an SDN controller with the slave role monitors only the system operation. In this situation any received Packet In message should be ignored by the SDN controller acting as slave. Finally, the SDN controllers with the equal role share among them the workload related to monitor and control the data forwarding. That is, the equal role suggests the same behavior when compared with the master role, but there are some important differences. Within a cluster of several SDN controllers only a single SDN controller can be selected as the master role. Thus, the remaining SDN controllers should act as slaves. Alternatively, when all the cluster controllers share the same role, i.e. they are equal, this implies that the data forwarding load should be shared among these controllers, but in a coordinated way. We discuss now a new cluster server (like ZooKeeper[1]) and a lightweight management protocol for the resilient and orchestrated operation of any number of SDN controllers. This cluster can manage any network topology, eventually formed by the aggregation of diverse networking domains. Without reducing the usage flexibility of the proposed solution, in Fig. 7 we present an illustrative scenario of the system control level, where are visible three entities, namely the cluster manager and two SDN controllers.

---

[1] Apache ZooKeeper, available at https://zookeeper.apache.org/ (verified in 29/07/2021)

Assuming the cluster manager was pre-configured to the Master/Slave mode, we now describe how this mode works. After the boot of each SDN controller, that controller selects a random integer (i.e. cont_id). Then, each controller sends the randomly selected number to the Cluster Manager (Fig. 7, message 1 or 2). The messages are sent via TCP sockets, where the manager and any controller have respectively the server role and client role.

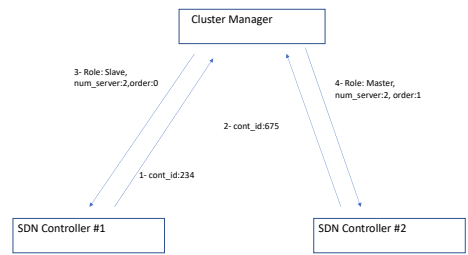

**FIGURE 7.** Deployed cluster-based architecture and its associated communication protocol that can manage any number of SDN controllers

After the cluster manager has received the initial messages from all the controllers previously configured to interact with the cluster manager, this manager evaluates three parameters that in the next protocol phase (messages 3 and 4) are returned to each controller. These three evaluated parameters are the role of each controller (in Fig. 7 controller #2 is the 'Master' because this controller obtained the highest cont_id), the total number of controllers (i.e. num_server) and the individual order of each controller (i.e. order). In respect to the last parameter, the cluster manager gives the order=0 to the controller that previously has reported the minimum cont_id value. In addition, the cluster manager gives the highest order value to the controller, which has previously reported the highest cont_id number.

We also need to have a distributed mechanism to establish an orchestrated management among the SDN controllers, avoiding conflicting decisions about what to do with the same Packet-In message simultaneously received by all those controllers. The expression (1) is used by each controller to decide (or not) on how to process the last received Packet-In message previously sent by an SDN-based switch with datapath identifier given by dpid. We should note that, as already explained, each controller has a unique order value.

$$\text{dpid mod}(num\_server) == order \quad (1)$$

When the equality in (1) becomes True, this occurs exclusively at a single controller among any set of controllers. Therefore, there is always a unique controller to decide how the message within the received Packet-In should be analyzed and processed. In this way, we have a distributed decision or consensus mechanism among the controllers. This solution offers the significant advantage of avoiding the exchange of signaling traffic directly between the controllers. Nevertheless, it has a significant drawback. It is not completely fair in terms of balancing the load among the controllers, i.e. taking an equal share, in scenarios where there are distinct amounts of data flows traversing the forwarding switches. In these situations, there is a distinct (perhaps unfair) workload level assigned to each controller. In the text below, we describe a fairer algorithm, which balances better the workload among controllers.

The alternative orchestration proposal is that each controller could decide if it processes or not any received Packet-In message as summarized in (2). The subtle difference in relation to (1) is the replace of dpid by packet_in_counter, which is the aggregated value of all received Packet-In messages by each controller. Assuming every switch is simultaneously connected via OpenFlow with every available controller, all the controllers share the same trend on the packet_in_counter parameter. The decision algorithm in (2) enables a fairer control load distribution among the diverse controllers, keeping also all the referred positives of the decision algorithm in (1).

$$\text{packet\_in\_counter mod}(num\_server) == order \quad (2)$$

The proposed solution based on the (random) identifier of SDN controllers can be evolved: the initial decision of the cluster manager based on cont_id can be later adjusted considering the control channel delay measurement between each controller and every switch. The posterior changes on controllers' orchestration made by the cluster manager can minimize the delay of each control channel. This will be studied in future work.

To finalize this sub-section, a final design observation is the need to guarantee status consistency within the SDN-based system after a switch has changed its controller. In this case, the new controller should replace the old flow rules stored in each controlled switch by new ones.

## IV. DEPLOYMENT

This section debates the deployment of several cross-level solutions for increasing the resilience of an SDN-based system in adverse scenarios such as system failures, high loads, or attacks. These deployment aspects have been aggregated in RECS, as explained at the beginning of section III. Due to the availability of multiple SDN controllers, there are also resilience gains at the switching logic system level.

### A. ENHANCING SYSTEM RESILIENCE AGAINST CONTROL CHANNEL COMMUNICATION FAILURE

The current sub-section presents the deployment of a solution that enables data forwarding devices to detect and react autonomously against communication failures to the intermediate system level, which is responsible for the implementation of the switching logic behind the system operation. As explained in sub-section III.A, the standalone functional mode is the more suitable option to support a higher degree of operational resilience at the data message forwarding level of our programmable edge system. In this way, we configured each OpenvSwitch to operate in the standalone mode, issuing the ovs-vsctl command visualized in Fig. 8.

```
# configure the switch to the standalone functional mode
sudo ovs-vsctl set-fail-mode s1 standalone
```
**FIGURE 8.** Configure OpenvSwitch s1 to standalone functional mode

### B. AVOIDING SERVER CONGESTION

To deal with the potential issue associated with server congestion in the network, we propose a solution based on a server farm, where several servers elastically offer the same service to a high number of clients. Our programmable solution deployed in the SDN controller has two main steps: i) manage the ARP protocol; and ii) manage the IP virtual address, which identifies the server farm. In the following, we explain the two algorithms behind these two steps.

#### 1) MANAGE THE ARP PROTOCOL

The several main processing steps of the SDN controller to manage the ARP protocol are summarized in Algorithm 1. Steps 1-9 of Algorithm 1 allow the SDN controller to identify an ARP message requesting the MAC address of the Server that should be associated with the IP virtual address (i.e. Virtual_IP in step 4). Then steps 10-15 select the MAC address of the Server that should be sent back to the client. The N parameter represents the number of available servers within the server farm. The number of available servers can be dynamically adjusted to the total client demand for the service provided by the edge server farm. The MAC address selected for each client follows a round-robin scheduling principle. This is an important characteristic of our solution to ensure the server load balancing among the clients. Steps 16-21 create the ARP Reply packet and return it to the initial calling code. In step 6 the ARP Reply packet is sent using a Packet-Out message to the switch, which by its turn sends the ARP Reply back to the client, following the instruction action of previous OpenFlow message. After receiving that packet, the client populates its ARP table with a new entry mapping Virtual_IP to the server returned MAC address.

**Algorithm 1: Manage the ARP protocol**

1:     **for each** Packet-In Event with pkt **do**
2:       **if** pkt.ether.type=ARP
3:         arp_header = pkt.get_protocol(arp)
4:         if arp_header.opcode == ARP_Request and arp_header.dst_ip == Virtual_IP
5:           ARP_reply_packet = generate_arp_reply(arp_header.src_ip,arp_header.src_mac)
6:           send_msg(Packet-Out(ARP_reply_packet))
7:         **end if**
8:       **end if**
9:     **end for**
10:   **generate_arp_reply**(dst_ip,dst_mac)

```
11:      arp_target_ip = dst_ip
12:      arp_target_mac = dst_mac
13:      src_ip = Virtual_IP
14:      i ← arp_target_mac mod(N)
15:      src_mac = MAC_addr[i]
16:      pkt = Packet()
17:      pkt.add_protocol(ethernet(dst=dst_mac,src=src_mac))
18:
         pkt.add_protocol(arp(opcode=ARP_Reply,src_mac=src_mac,src_i
         p=src_ip,dst_mac=arp_target_mac, dst_ip=arp_target_ip))
19:      pkt.serialize()
20:      return pkt
21:   end function
```

### 2) MANAGE THE IP VIRTUAL ADDRESS

After the MAC address of the selected server has been returned to the client, the same client sends a packet destined to the Virtual_IP. Then, the packet IP destination address should be changed to the IP address used by the selected server. This change is made in the last switch on the destination path before the packet arriving to the selected server. Otherwise, the selected server will not reply. The Algorithm 2 summarizes how the IP address is changed.

**Algorithm 2: Manage the IP virtual address**
```
1:    for each Packet-In Event with pkt do
2:       datapath = Event.msg.datapath
3:       parser = datapath.ofproto_parser
4:       in_port = msg.match['in_port']
5:       eth = pkt.get_protocol(ethernet)
6:       src_mac = eth.src
7:       dst_mac = eth.dst
8:       if eth.type==IPv4 and datapath.id == last_switch_before_Server
9:          ip_header = pkt.get_protocol(ipv4)
10:         up_header = None
11:         if ip_header.proto == TCP
12:            up_header = pkt.get_protocol(tcp)
13:         elif ip_header.proto == UDP
14:            up_header = pkt.get_protocol(udp)
15:         elif ip_header.proto == ICMP
16:            up_header = pkt.get_protocol(icmp)
17:         else
18:            print "Unsupported protocol!"
19:         end if
20:
            handle_ip_packet(datapath,in_port,ip_header,up_header,pars
            er,dst_mac,src_mac)
21:      end if
22:   end for
23:   handle_ip_packet(datapath,in_port,ip_header,up_header,parser,dst
      _mac,src_mac)
24:      dpid=datapath.id
25:      if ip_header == virtual_IP
26:         for each i in all Servers do
27:            if dst_mac == Server_MAC[i]
28:               server_dst_ip = Server_IP[i]
29:               server_out_port = Server_Port[i]
30:            end if
31:         end for
32:         path = networkx.shortest_path(net,dpid,dst_mac)
33:         next = path[1]
34:         out_port = net[dpid][next]['port']
35:         match =
            parser.OFPMatch(in_port=in_port,eth_type=IPv4,ipv4_src=ip
            _header.src,ipv4_dst=Virtual_IP)
36:         actions =
            [parser.OFPActionSetField(ipv4_dst=server_dst_ip),parser.OF
            PActionOutput(out_port)]
37:         add_flow(datapath,match,actions)
38:         match =
            parser.OFPMatch(in_port=server_out_port,eth_type=IPv4,ipv
             4_src=server_dst_ip,ipv4_dst=ip_header.src)
39:         actions =
            [parser.OFPActionSetField(ipv4_src=ip_header.dst),parser.OF
            PActionOutput(in_port)]
40:         add_flow(datapath,match,actions)
41:         generate_modified_packet()
42:      end if
43:   end function
```

Steps 1-19 of Algorithm 2 allow the SDN controller to discover in the received IP message which protocol is being used at the layer above the network layer. The eventual change of the IP address is made inside the function handle_ip_packet() (step 20). Inside this function (steps 26-31), the SDN controller selects the server IP address (step 28) according to the destination MAC address sent by the client. This last MAC address was learned by the client via the ARP protocol (see Algorithm 1). In addition, executing step 29, the SDN controller selects the output_port of the switch directly attached to the network interface of the Server. Steps 32-34 find the shortest path between the current switch and destination server, including the output switch port of the forwarding path. Then, the SDN controller specifies to the current switch a list of two actions (step 36): i) change the destination IP address to the one being used by the selected Server network interface; ii) specify the output previously found in step 34. Step 37 sends to the switch a flow rule to be applied to the future packets of the same flow. Steps 38-40 are necessary to create the correct flow rule for the reverse path (i.e. traffic from the server to the client) and send that rule to the switch to be applied to the future packets of the same flow but in the reverse direction. Finally, step 41 is needed for the SDN controller to make a copy of the first received packet of the flow being processed but making the necessary adjustment in the IP address and sending back that changed packet to the switch. The SDN controller has also the capability of detecting the need to automatically activate some part of the network edge topology before sending a packet destined to any node within that topology part.

### C. OVERCOME LINK CONGESTION AND PROTECT TCP QUALITY AGAINST UDP RESOURCE USAGE UNFAIRNESS

This sub-section deals with the deployment at the topological switches of several Select groups (i.e. one for each server of a server farm). The usage of these Select groups offer two pertinent system functional advantages: balancing the traffic load and mitigating the negative impact of any eventual loop in the traffic forwarding through the network topology. Despite the usage of the Select groups, we have implemented a specific mechanism at the SDN controller to eliminate any eventual loop at the data message forwarding. This mechanism is presented below as Algorithm 3.

**Algorithm 3: Avoid any found loop at the data message forwarding**
```
1:    for each Packet-In Event with pkt do
2:       datapath = Event.msg.datapath
```

```
3:      dpid = datapath.id
4:      in_port = msg.match['in_port']
5:      eth = pkt.get_protocol(ethernet)
6:      src_mac = eth.src
7:      dst_mac = eth.dst
8:      if src_mac not in self.learned_macs[dpid]:
9:          self.learned_macs[dpid][src_mac] = in_port
10:     else:
11:         if in_port != self.learned_macs[dpid][src_mac] and dst_mac
            == 'ff:ff:ff:ff:ff:ff':
12:             return
13:         endif
14:     endif
15: end for
```

In steps 8-9, the SDN controller learns a first seen frame received at the switch identified by dpid, with a specific MAC source address and received in_port of that switch. In the case of a future event associated to a frame from the same switch with repeated physical source and destination addresses but with a distinct received switch port from the previously learned port (step 11), the SDN controller classifies this more recent event as induced by a topology loop. Then, the controller stops the processing of Packet-In handler function (step 12). In this way, there is no Packet-Out being sent to the switch and the loop situation is cancelled.

In the text below, we will explain the diverse algorithms to deploy a set of relevant controller functionalities for supporting routing decisions based on the shortest path to each destination. These controller functionalities are summarized in Fig. 9 and are as follows: i) periodic retrieval of operational statistics from switches; ii) update of link costs and Select buckets weights; iii) send back updated Select groups to switches.

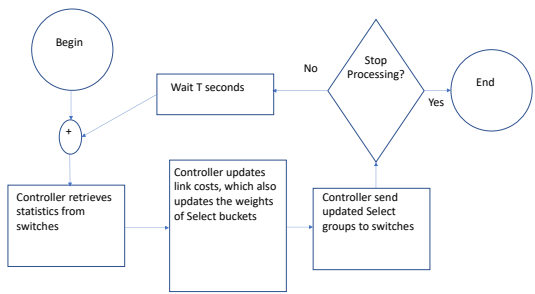

**FIGURE 9.** Control loop that retrieves statistical information from switches, enables the controller to update link costs and the weights of Select buckets, which follow the parallel cost paths to each specific destination, and then ends with the controller sending renewed Select groups to the switches. These groups allow load balancing at the data forwarding level. The loop reruns periodically (once every T seconds)

### 1) CONTROLLER PERIODICALLY RETRIEVES DATA FORWARDING STATISTICS

The major processing events of the SDN controller to retrieve periodically port and flow rule statistics from all the topological switches are summarized below as Algorithm 4. To run the statistics collection, we use a thread that is launched in step 2. As already said, this thread runs once every T second, which is possible due to steps 5, 10 and 11. The statistics collection from all the switches is initiated in steps 6-8. Further details about the start of this data forwarding status collection are available in steps 13-20.

**Algorithm 4: The controller retrieves periodically statistics from the data forwarding level (once each T second)**
```
1:  def __init__(self, *args, **kwargs):
2:      self.monitor_thread = hub.spawn(self._monitor)
3:  end function
4:  def _monitor(self):
5:      while True do
6:          for each id in self.switches do
7:              datapath = self.datapath_list[id]
8:              self._request_stats(datapath)
9:          end for
10:         hub.sleep(T)
11:     end while
12: end function
13: def _request_stats(self, datapath):
14:     ofproto = datapath.ofproto
15:     parser = datapath.ofproto_parser
16:     req = parser.OFPFlowStatsRequest(datapath)
17:     datapath.send_msg(req)
18:     req = parser.OFPPortStatsRequest(datapath, 0,
        ofproto.OFPP_ANY)
19:     datapath.send_msg(req)
20: end function
```

### 2) CONTROLLER UPDATES LINK COSTS

The more relevant processing steps of the SDN controller to obtain switch port statistics and after updating the links costs using some of those statistics are summarized below as Algorithm 5. We use here a Python decorator (step 1) specialized in a very specific event type, which is generated within the environment of the SDN controller, after the controller has received an OpenFlow message from the switch transporting the statistical data from all the available ports of that switch. In step 2 we declare the function which is called each time the SDN controller receives a port statistics message from any switch. The processing of the received statistical data and consequent update of each link cost are respectively in steps 5-17 and step 18. In the last processing, we have used exponential moving average with alfa (steps 7 and 9) assuming the value of 0.2. This value has been experimentally selected from the range [0, 1].

**Algorithm 5: The controller gets port statistics sent by switches and updates link costs using some selected statistical data**
```
1:  @set_ev_cls(ofp_event.EventOFPPortStatsReply,
    MAIN_DISPATCHER)
2:  def _port_stats_reply_handler(self, ev):
3:      body = ev.msg.body
4:      dpid = ev.msg.datapath.id
5:      for each stat in body do
6:          port = stat.port_no
7:          self.rate[dpid][port] = alfa * self.rate[dpid][port] + (1-alfa) *
            (stat.tx_bytes - self.tx_bytes[dpid][port])
8:          self.tx_bytes[dpid][port] = stat.tx_bytes
9:          self.bandwidths[dpid][port]     =             alfa    *
            self.bandwidths[dpid][port] + (1-alfa) * self.rate[dpid][port]
10:     end for
11:     for each port_sw in range (6) do
```

```
12:        c = 1 - (self.DEFAULT_BW1 / ( k *
           self.bandwidths[dpid][port_sw]))
13:        if c <= 0:
14:           cost = 1000
15:        else:
16:           cost = 10 / c
17:        end if
18:        self.costs[dpid][port_sw] = cost
19:     end for
20:  end function
```

In Algorithm 6, we briefly present the major processing steps of SDN controller to update link weights of the NetworkX algebraic topology with the costs evaluated as already explained in Algorithm 5. The cost links enable the controller to take routing decisions based on the shortest path to each destination. We have opted to deploy this processing in an event-triggered way instead of using periodical processing, because the former option consumes less resources from the controller than the latter alternative.

**Algorithm 6: The controller updates the link weights of the NetworkX algebraic topology with the costs evaluated @ Algorithm 5**

```
1:  for each Packet-In Event with pkt do
2:     for each node ni in topology do
3:        for each direct neighbor of node ni in topology do
5:           self.net.add_weighted_edges_from([ (ni,direct neighbor of
              node ni,self.costs[ni][output port from ni to direct neighbor
              of node ni]), ])
6:        end for
7:     end for
8:  end for
```

### 3) CONTROLLER UPDATES THE BUCKET WEIGHTS OF SELECT GROUPS

Algorithm 7 describes how the SDN controller updates the bucket weights of the Select groups enabled within the SDN-based system. The bucket weights directly reflect the cost paths. Each cost path is the sum of the diverse link costs forming that path. The SDN controller only updates in an event-triggered way the bucket weights of a Select group, after receiving a Packet-In message (step 1) that controller is responsible to control. We again opt for this, despite performing it periodically, to save controller processing resources. In steps 17-24 the switches Select groups are installed or updated.

**Algorithm 7: Controller updates bucket weights of Select groups**

```
1:   for each Packet-In Event with pkt do
2:      datapath = Event.msg.datapath
3:      dpid = datapath.id
4:      in_port = msg.match['in_port']
5:      eth = pkt.get_protocol(ethernet)
6:      src_mac = eth.src
7:      dst_mac = eth.dst
8:      if dst_mac in self.mac_to_port[dpid]:
9:         out_port = self.mac_to_port[dpid][dst_mac]
10:     else:
11:        out_port = ofproto.OFPP_FLOOD
12:     end if
13:     if out_port != ofproto.OFPP_FLOOD:
14:        self.install_paths(datapath, in_port, src_mac, dst_mac)
15:     end if
16:  end for
17:  def install_paths(self, dp, in_port, src, dst):
18:     obtain from network algebraic topology all possible paths
        between current switch (dp.id) and the destination (dst)
19:     if more than one path to dst:
20:        create and install (or update) a Select group with a bucket for
           each possible path; each bucket has a weight equal to the
           cost of the associated path
21:     else:
22:        create and install (or update) flow rules for the single next-
           hop for the destination
23:     end if
24:  end function
```

### D. ORCHESTRATING A CLUSTER OF SDN CONTROLLERS AND THE CONTROL CHANNEL LOAD

We aim to develop a new cluster server, like Apache ZooKeeper, but using a new lightweight management protocol for the resilient operation of a programmable system supported by multiple controllers. Algorithm 8 describes the processing steps of a multi-thread cluster server that manages any number of SDN controllers sharing the same cluster. The multi-thread implementation for the cluster server offers some performance and scalability gains. Thus, each SDN controller only needs to establish an initial and single TCP connection with the cluster manager and keep it active during the time the SDN controller is running. Consequently, we diminish the network overload in relation to the simpler implementation of a common thread to deploy the cluster manager for all the SDN controllers. System gains in terms of scalability and performance of the multi-thread cluster manager become more relevant as there are many more SDN controllers. In steps 28-33, the cluster server finds out the SDN controller with the highest communicated identifier. This SDN controller is selected by the cluster server as the MASTER of the cluster and all other SDN controllers are selected as SLAVEs. The relevant decisions are made when the parameter self.EQUAL (step 9; steps 35-40) is False. Otherwise (step 41), all the SDN controllers assume the EQUAL role. In this situation, we need to deploy a correct orchestration mechanism among all controllers to avoid two or more of them simultaneously controlling either a switch or Packet-In message from the data forwarding system level. To support this distributed orchestration mechanism, the server cluster (step 42) sends to each SDN controller the EQUAL role and, the order and count parameters. The count parameter is the total number of SDN controllers; order enables distributed orchestration (see III.D) between any number of controllers.

Algorithm 9 presents the SDN controller operation when it checks with the cluster manager which role that SDN controller should assume (steps 6-15). If the SDN controller is alone in the control cluster, then that SDN controller should assume the MASTER role (steps 13-15). Otherwise (steps 8-11), the SDN controller should assume the role reported by the cluster manager, the number of SDN controllers and the order number assigned by the same cluster manager to each SDN controller. In addition, as the SDN controller detects a role change (step 16), then it needs to inform all the switches about that change (step 17) and, if necessary, the SDN controller should delete all the old flow rules from the switches which are now under its control before installing new flow rules (step 18). The update of flow rules is important to ensure a coherent and reliable control of the data forwarding system level.

**Algorithm 8: The multi-thread cluster manager scalably manages any number of controllers belonging to the same control cluster**

```
1:    class ThreadedClusterServer(object):
2:      def __init__(self, host, port):
3:        self.host = host
4:        self.port = port
5:        self.sock = socket.socket(socket.AF_INET, socket.SOCK_STREAM)
6:        self.sock.setsockopt(socket.SOL_SOCKET, socket.SO_REUSEADDR, 1)
7:        self.sock.bind((self.host, self.port))
8:        self.mydict = {}
9:        self.EQUAL = False
10:     end function
11:     def listen(self):
12:       self.sock.listen(5)
13:       while True do
14:         client, address = self.sock.accept()
15:         client.settimeout(60)
16:         threading.Thread(target = self.listenToClient,args = (client,address)).start()
17:       end while
18:     end function
19:     def listenToClient(self, client, address):
20:       size = 1024
21:       while True do
22:         try:
23:           data = client.recv(size)
24:           if data:
25:             self.mydict[data] = 1
26:             max=0
27:             count = []
28:             for key in self.mydict do
29:               count.append(key)
30:               if max < key:
31:                 max = key
32:               end if
33:             end for
34:             order = count.index(data)
35:             if not self.EQUAL:
36:               if data == max:
37:                 client.sendall("MASTER" + ":" + str(len(count)) + ":" + str(order))
38:               else:
39:                 client.sendall("SLAVE" + ":" + str(len(count)) + ":" + str(order))
40:               end if
41:             else:
42:               client.sendall("EQUAL" + ":" + str(len(count)) + ":" + str(order))
43:             end if
44:           else:
45:             raise error('Client disconnected')
46:           end if
47:         except:
48:           self.mydict = {}
49:           client.close()
50:           return False
51:       end while
52:     end function
53:   if __name__ == "__main__":
54:     while True do
55:       port_num = input("Port? ")
56:       try:
57:         port_num = int(port_num)
58:         break
59:       except ValueError:
60:         pass
61:     end while
62:     ThreadedClusterServer('',port_num).listen()
```

**Algorithm 9: Each controller communicates periodically with the top level cluster manager (once each T second)**

```
1:    def __init__(self, *args, **kwargs):
2:      self.monitor_thread = hub.spawn(self._monitor)
3:    end function
4:    def _monitor(self):
5:      while True do
6:        resp=self.check_role()
7:        list = resp.split(":")
8:        if len(list) > 1:
9:          self.mode = list[0]
10:         self.num_serv = list[1]
11:         self.order = list[2]
12:       end if
13:       if self.num_serv == '1':
14:         self.mode = 'MASTER'
15:       end if
16:       if self.mode != self.mode_prev:
17:         self.send_role_request(for all switches)
18:         self.load_default_rules(for all switches under the control of this controller)
19:       end if
20:       self.mode_prev = self.mode
21:       hub.sleep(T)
22:     end while
23:   end function
```

**Algorithm 10: Each controller assumes the role EQUAL avoiding any conflict with other controllers**

```
1:    for each Packet-In Event with pkt do
2:      datapath = Event.msg.datapath
3:      dpid = datapath.id
4:      if self.mode == 'EQUAL':
5:        if not (dpid % int(self.num_serv) == int(self.order)):
6:          return
7:        else:
8:          Analyse, process and control the current message
9:        endif
10:     endif
11:   end for
```

Algorithm 10 shows one possible way to support the coordinated control among all the SDN controllers sharing the same role EQUAL, avoiding potential conflicts among them (steps 5-6). Our solution is easily modified to support other options to coordinate the SDN controllers, including a distinct scenario such as MASTER-SLAVEs. We have already discussed some alternative orchestration methods for the scenario where all the SDN controllers assume the EQUAL role in sub-section III.D.

Table II summarizes the ten algorithms discussed in Section IV, indicating for each algorithm its specific goal and broader proposal aim. These algorithms can be divided into three groups. The first group is composed by algorithms 1-2, which aim to avoid congestion at the dataplane servers. These edge servers are made accessible to clients via IP virtual address, and the number of active servers depends on the number of flows requiring the common service offered by those virtualized servers. The second group is formed by algorithms 3-7 with the intent of avoiding congestion at data forwarding links by using load balancing techniques. The TCP quality protection against UDP unfair resource competition is also achieved. The third, last, group contains algorithms 8-10 that target to orchestrate with low network overhead a set of redundant SDN controllers at the switching logic level.

TABLE II. ALGORITHMS SUMMARIZATION AND THEIR GOALS

| Algorithm | Sub-Section | Algorithm goal | Proposal goal |
|---|---|---|---|
| 1 | IV.B | Manage ARP protocol | Avoids server congestion |
| 2 | IV.B | Manage the IP virtual address | |
| 3 | IV.C | Avoid any topology loop | Overcome link congestion and protect TCP quality against UDP resource usage unfairness |
| 4 | IV.C | Controller periodically retrieves data forwarding statistics | |
| 5 | IV.C | Controller updates link costs | |
| 6 | IV.C | Controller updates the edge weights of the NetworkX algebraic topology evaluated by alg. #5 | |
| 7 | IV.C | Controller updates the bucket weights of Select groups | |
| 8 | IV.D | Multi-thread cluster manager that manages a set of controllers | Distributed orchestration of any number of redundant SDN controllers |
| 9 | IV.D | Each controller communicates periodically with the top level cluster manager | |
| 10 | IV.D | Each controller assumes the role EQUAL avoiding any functional conflict with other controllers | |

## V. EVALUATION

Now we discuss the evaluation of RECS, our proposal for increasing the SDN-based system resilience against real-life network issues that could penalize its performance. Fig. 10 visualizes the RECS SDN-based system under evaluation.

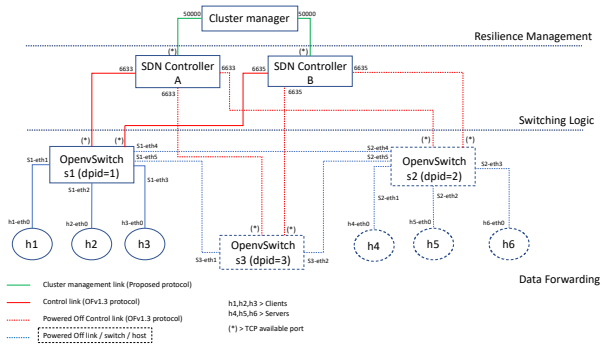

**FIGURE 10.** RECS - a SDN-Based system under evaluation. This system aims to beneficially manage the runtime adaptation of computational resources from edge cloud systems to mitigate against disruptive events on these systems

There are only two SDN controllers, but our proposal is capable of scaling for the number of supported SDN controllers. We are using a distributed and flat programmable switching logic, which is orchestrated by a cluster manager. Some data forwarding entities are initially powered off, as visualized in Fig. 10 as "Powered Off"; the elastic resource activation is omitted. Thus, the initial minimalist network topology assumes a strong objective, to minimize energy consumption by switching off network sectors that are not required for normal network operation. Then, the network domain can evolve to an operational state where an increasing number of new data flows need to be controlled in terms of their destination-based routing. As follows, the network devices that receive the new flows use OpenFlow messages to inform the control plane about the need for more (virtualized) computational and networking resources to fulfill the requisites of those flows; these extra resources are promptly activated by the control plane before forwarding actions are sent back to data plane switching devices.

To enable the reproducibility of the results of this work, Table III summarizes the hardware and software used during our tests. The tests made to evaluate RECS system are summarized in Table IV. Also, the main objective of each test is identified.

TABLE III. HARDWARE AND SOFTWARE TOOLS USED DURING THE EVALUATION TESTS

| ASUS Intel® Core™ i7-3517U CPU @ 1.90GHz 2.40GHz, 12 GB RAM, Windows10 Education x64 | - |
|---|---|
| VirtualBox Ubuntu 18.04 | https://www.virtualbox.org/; https://releases.ubuntu.com/18.04/ |
| Ryu SDN Controller (v4.15) | https://ryu-sdn.org/ |
| NetworkX (v1.11) | https://networkx.org/ |
| OpenvSwitch (v2.9.8, DB Schema 7.15.1) | https://www.openvswitch.org/ |
| Python 2.7.17 | https://www.python.org/download/releases/2.7/ |
| ip utility, iproute2-ss180129 | https://git.kernel.org/pub/scm/network/iproute2/iproute2.git |
| iperf3, iperf2 | https://github.com/esnet/iperf, https://iperf2.sourceforge.io/iperf-manpage.html |
| Wireshark | https://www.wireshark.org/ |

TABLE IV. EVALUATION TESTS

| Section | Main aspect(s) under analysis |
|---|---|
| V.A | Switch operational resilience after being disconnected from the upper level programmable switching logic |
| V.B | Mitigation of congestion at topological server network interface |
| V.C | Mitigation of congestion at topological switch ports |
| V.D | Orchestrating a cluster of SDN controllers and their workload |

### A. ENHANCING SYSTEM RESILIENCE AGAINST CONTROL CHANNEL COMMUNICATION FAILURE

We have tested the standalone mode (see III.A) of a software-based switch such as the case of OpenvSwitch. It was studied the scenario of how the system behaves after the switches become disconnected from the SDN controllers. In our testbed, we have tried to ping between hosts in two distinct scenarios. In the first scenario, the switch can communicate with the SDN controller. The ping results of this test are available in Fig. 11.

```
sudo ip netns exec h1 ping -c4 10.0.0.2
PING 10.0.0.2 (10.0.0.2) 56(84) bytes of data.
64 bytes from 10.0.0.2: icmp_seq=1 ttl=64 time=18.5 ms
64 bytes from 10.0.0.2: icmp_seq=2 ttl=64 time=0.188 ms
64 bytes from 10.0.0.2: icmp_seq=3 ttl=64 time=0.218 ms
64 bytes from 10.0.0.2: icmp_seq=4 ttl=64 time=0.155 ms

--- 10.0.0.2 ping statistics ---
4 packets transmitted, 4 received, 0% packet loss, time 3040ms
rtt min/avg/max/mdev = 0.155/4.783/18.571/7.960 ms
```

**FIGURE 11.** Ping test before the communication failure between SDN controllers and the switches

Analyzing and comparing the RTT of the first ping tentative against the RTTs of the next tries of the same ping command, we can notice a significantly higher RTT value (i.e. 18.5ms in Fig. 11) for the first tentative when compared with the following tentatives (i.e. within the range [0.155, 0.218]ms). This RTT difference is due to the fact the first ping try is controlled in a reactive way, involving the SDN controller in the final decision about how to route the ICMP messages through the network topology[2]. In addition, the ping tries following the first one are already proactively controlled. This means that at the time the messages of those ping attempts arrive at each switch on the path to the destination of each ICMP message, that switch has already local flow rules for commuting directly the ICMP traffic without involving anymore the SDN controllers.

In the second scenario, we have stopped all the SDN controllers and then executed again the ping command of the first scenario. The ping results of the second scenario are presented below in Fig. 12. Analyzing these results, we can conclude that the diverse ping tentatives have similar and low values in their RTTs. In addition, we have not detected any failure. In this way, we can conclude the switches after becoming disconnected from the SDN controllers continue their expected operation as Link Layer Learning switches.

```
sudo ip netns exec h1 ping -c4 10.0.0.2
PING 10.0.0.2 (10.0.0.2) 56(84) bytes of data.
64 bytes from 10.0.0.2: icmp_seq=1 ttl=64 time=0.580 ms
64 bytes from 10.0.0.2: icmp_seq=2 ttl=64 time=0.194 ms
64 bytes from 10.0.0.2: icmp_seq=3 ttl=64 time=0.212 ms
64 bytes from 10.0.0.2: icmp_seq=4 ttl=64 time=0.108 ms

--- 10.0.0.2 ping statistics ---
4 packets transmitted, 4 received, 0% packet loss, time 3061ms
rtt min/avg/max/mdev = 0.108/0.273/0.580/0.182 ms
```

**FIGURE 12.** Ping test after the communication failure between SDN controllers and the switches.

### B. AVOIDING SERVER CONGESTION

This sub-section evaluates the proposed mechanism to balance the load of a high number of clients among a dynamic pool of servers of a server farm (or serverless edge computing cluster). It presents and discusses the performance results of two tests. In the first test, we aim to measure the activation time of a set of devices of the network topology, including the virtualized services, before transferring some traffic through that activated topology part towards the virtualized edge services. The performance results of five ping tries sent from h1 client towards the server farm IP Virtual Address are presented in Fig. 13.

```
ip netns exec h1 ping -c5 10.0.0.10
PING 10.0.0.10 (10.0.0.10) 56(84) bytes of data.
64 bytes from 10.0.0.10: icmp_seq=1 ttl=64 time=2648 ms
64 bytes from 10.0.0.10: icmp_seq=2 ttl=64 time=1630 ms
64 bytes from 10.0.0.10: icmp_seq=3 ttl=64 time=607 ms
64 bytes from 10.0.0.10: icmp_seq=4 ttl=64 time=0.484 ms
64 bytes from 10.0.0.10: icmp_seq=5 ttl=64 time=0.166 ms

--- 10.0.0.10 ping statistics ---
5 packets transmitted, 5 received, 0% packet loss, time 4058ms
rtt min/avg/max/mdev = 0.166/977.446/2648.303/1026.179 ms, pipe 3
```

**FIGURE 13.** RTT trend of ICMP traffic towards a destination virtualized node which in the beginning is not running and it should be automatically powered on before receiving and processing the traffic data

In the current test, the activated devices were two OpenvSwitch devices, three server nodes, and five necessary links to interconnect all those virtualized devices (see Fig. 10). Analyzing the RTT of the diverse ping tries, one can conclude that the first ping tentative has suffered the highest RTT (i.e. 2.65s). This highest value is due to several reasons: i) the ARP protocol and the associated processing of server MAC address (see sub-sections III.B and IV.B for further information on this); ii) the reactive control of both the ICMP request and ICMP reply of the first ping tentative; and last and not least, iii) the time to power on the virtualized network infrastructure and edge servers. Considering a time interval of 1s between two consecutive ping attempts for the same destination, the additional delay imposed by all the three previous referred aspects is gradually being reduced as one can notice in the RTT of pings #2 and #3. The RTT of ping #4 indicates the associated ICMP messages were already autonomously controlled in each switch on the path to the selected destination of the server farm.

The second and last test of the current sub-section studies the system performance when a server farm delivers a common service to many clients. This common service is transported over TCP. To produce and consume the TCP flows we have used some iperf3 commands as shown in Table V.

**TABLE V.** IPERF3 COMMANDS

|  | Iperf3 command | Main objectives |
|---|---|---|
| **Server** | ip netns exec h4 iperf3 -p 5002 -s | Server is waiting for client requests at TCP port 5002 |
| **Client** | ip netns exec h1 iperf3 -f m -c 10.0.0.10 -p 5002 -t 33 -i 33 -l 1448 -M 1460 -O 3 -T 0 -P #TCP_flows \| grep SUM >> $TMPFILE | Host h1 sends TCP segments to the cluster IP address through concurrent TCP connections; the entire TCP traffic originated in h1 is diverted to a single server host (h4) by the SDN system |

The results of the test are shown in Fig. 14. They have been obtained from five trials for each number of concurrent TCP flows. From these results, one can conclude that the usage of a server farm by itself does not offer a significant quality improvement on the service provided to the clients, because all the TCP flows destined to a single server follow the same network path, which offers constrained connectivity resources (i.e. all the topology links are rate limited to 10 Mbps by the Linux traffic control - tc tool) to the aggregated traffic load.

---

[2] The largest RTT for the initial ping is also due to the extra delay imposed by the ARP protocol.

For the current testing scenario, the usage of a server farm is not enough by itself to ensure the aimed quality of service for each TCP flow. The next sub-section evaluates a possible solution to ameliorate the benefits on the service quality provided by the server farm.

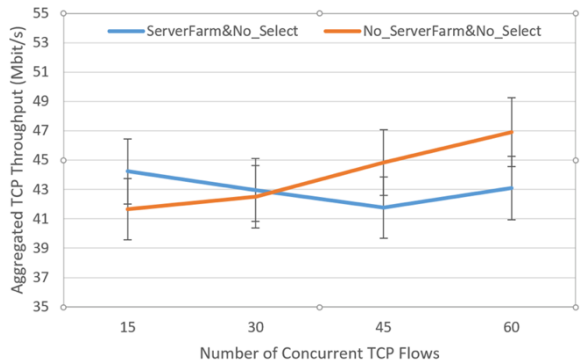

**FIGURE 14.** Performance comparison between the single server scenario and the server farm instance

### C. OVERCOME LINK CONGESTION AND PROTECT TCP QUALITY AGAINST UDP RESOURCE USAGE UNFAIRNESS

From the results of sub-section V.B, we have concluded that the Quality of Service provided by a server farm should be enhanced. Aligned with this goal, we have planned to enable in our SDN-based system a routing load balancing mechanism supported by Select groups at the network switches. In this way, we have repeated the test presented at the end of last sub-section but now with the usage of Select groups at the data message forwarding. The results of this test are visualized in Fig. 15.

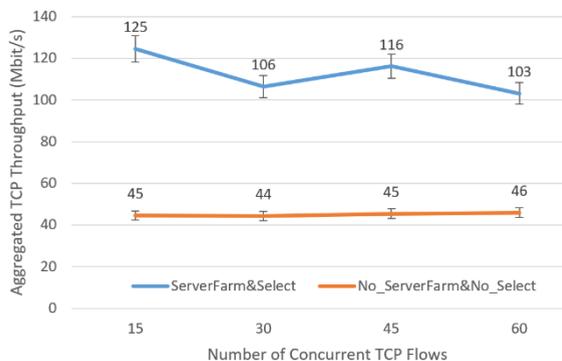

**FIGURE 15.** TCP Performance comparison at the server side between scenarios single server vs. server farm with Select groups being only used for the last scenario

The results of Fig. 15 have been obtained from five trials for each number of concurrent TCP flows. Comparing the results of the two scenarios under test, one can conclude that for the case the server farm and load balancing of Select groups are both enabled, the associated TCP aggregated throughput is roughly 178% (i.e. (125 - 45) * 100 / 45) higher than the throughput verified in the scenario where the Select groups are not active. In the current scenario, the server farm can offer an enhanced service to their clients when the corresponding TCP traffic is conveniently balanced among the multiple access links of the server farm by using Select groups in the switches belonging to the data forwarding level.

Another important advantage to be obtained from using Select groups is to deter network link congestion when messages using distinct transport protocols are competing for the same network resources. From Fig. 16, when Select groups are not used, the TCP traffic performance is penalized by 73% (i.e. (131-36)/131*100) due to the unfairness competition of UDP traffic (i.e. 3 x 2.2 Mbps) and the absence of any load balancing mechanism to countermeasure the network congestion strongly induced by UDP flows.

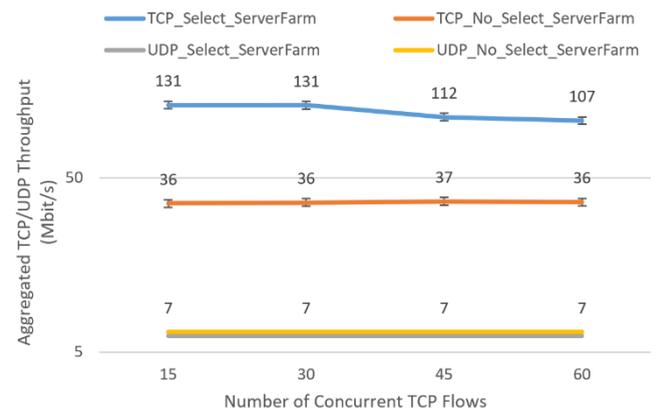

**FIGURE 16.** TCP performance comparison at the server side between scenarios with and without Select groups, when there are always three concurrent UDP flows with an individual constant source bit rate of 2.16Mbps (note: the vertical axis is presented in a logarithmic scale)

These results suggest Select groups are also relevant to protect the performance of TCP traffic in case there are competitive flows that monopolize the network resources.

### D. ORCHESTRATING A CLUSTER OF SDN CONTROLLERS AND THE CONTROL CHANNEL LOAD

This sub-section evaluates the two methods discussed in III.D for enabling the coordination among SDN controllers. Comparing these methods (Fig. 17), the orchestration method identified as III-D:Exp. (2) is the fairest one in terms of balancing the workload among the SDN controllers. This fairness enhancement occurs because III-D:Exp. (2) applies round robin scheduling to Packet In (PI) control channel event counter and all the controllers count the same PIs.

Nevertheless, as shown in Fig. 18, the orchestration method III-D:Exp. (2) shows a slightly higher channel control load when compared with the orchestration option III-D:Exp. (1). This difference on the channel control load between the two methods is due to the fairest method forcing each switch to be controlled by either one of the two SDN controllers, which implies more control messages in the control channel for example to indicate to each switch the new SDN controller. Alternatively, the method which is more unfair in terms of balancing the workload of SDN controllers allows each SDN controller to control the same set of switches during the entire working session.

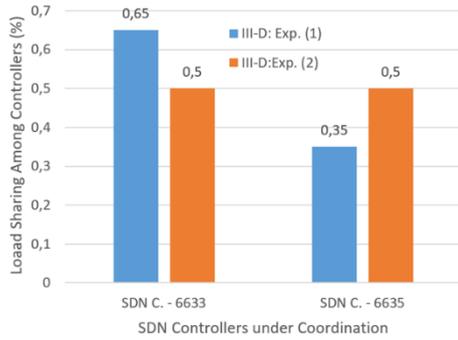

**FIGURE 17.** Channel control load sharing among controllers for the two distributed coordination mechanisms under comparison

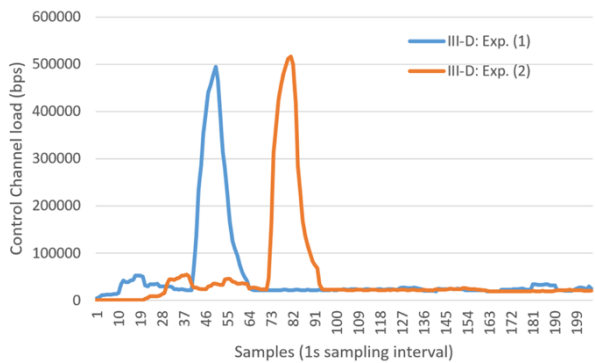

**FIGURE 18.** Channel control load of the two distributed coordination mechanisms during the same simulated avalanche of traffic at the data forwarding level

Thus, it seems there is a trade-off between increasing the fairness level in how the workload of the SDN controllers is balanced and the additional load on the control channel to support that increase on the fairness level.

Considering the scenario when the fairest coordination mechanism was used, we have also investigated the principal causes for the channel control (and cluster management as well because both share the same loopback interface for exchanging messages) peak load value. The results of this analysis are visualized in Fig. 19.

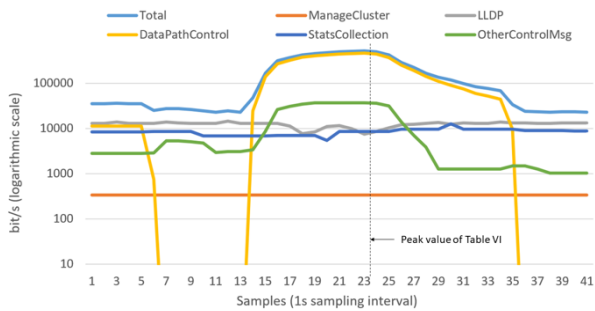

**FIGURE 19.** The decomposition of the fairest orchestration method channel control load among the diverse types of control/management messages

For the peak value time of Fig. 19, we show in Table VI how that value can be decomposed in the diverse types of control/management messages. Comparing the network overload induced by both control and management messages, we can make two main observations. First, the control messages (i.e. Packet-In, Packet-Out, Flow-Mod) exchanged between the SDN controllers and the data forwarding switches are responsible for 90% of the peak value of Fig. 19. The second observation is that the messages related to the management of the cluster of SDN controllers are only marginally responsible, with 0.1% for the same peak.

**TABLE VI.** DECOMPOSITION OF PEAK VALUE FROM FIGURE 19

| Messages | bit/s | Percentage (%) |
|---|---|---|
| Total | 516 868.6 | 100.0 |
| ManageCluster | 335.2 | 0.1 |
| LLDP | 7 481.6 | 1.4 |
| DataPathControl | 464 118.4 | 89.8 |
| StatsCollection | 8 659.2 | 1.7 |
| OtherControlMsg | 36 608.8 | 7.1 |
| Note: LLDP – Link Layer Discovery Protocol | | |

Then, we have compared the performance of two different implementations for the cluster manager. The first implementation can attend each time a single SDN controller, implying the cluster manager to disconnect the TCP connection with the last SDN controller before connecting to the next SDN controller. The second deployment of the cluster manager has a multi-thread design. Therefore, each SDN controller can keep active the TCP connection with the cluster manager, in parallel with other SDN controllers, during the entire working session. Analyzing the results visualized in Fig. 20, the scenario using a multi-thread design in the cluster manager significantly diminishes the load increase induced by the signaling traffic exchanged between the cluster manager and the diverse SDN controllers sharing the same control cluster in relation to the other option based on a single-thread design. Here, we have a trade-off between the complexity level of the cluster server implementation and the overload level on the management channel. In addition, Fig. 21 illustrates that the higher level of complexity associated to the multi-thread implementation of the cluster manager implies the allocation of a higher amount of processing and memory resources to run that implementation than the single-thread alternative.

After the master SDN controller fails, the new controller does not immediately control the system. To evaluate the efficiency of RECS controller failover mechanism, we have measured the failover time. In this experiment, a software switch connects two hosts that continuously exchange ICMP packets with a time interval of 1ms between two consecutive tries. We bring down the master, wait for the slave to become the new master, and activate again the previous failed controller, which could assume the role of slave or master, depending on its own novel id number. The obtained results have evidenced no ping failure during the state changes at control level. Consequently, the system failover time is lower than 1ms.

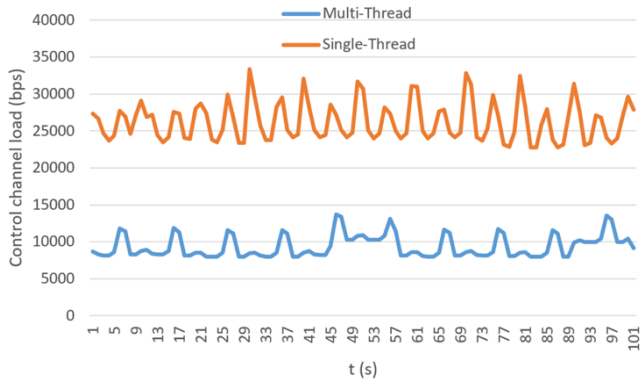

**FIGURE 20.** The network overload induced by the exchange of signaling traffic between the SDN controllers and the cluster manager for the two distinct implementations of that manager under comparison

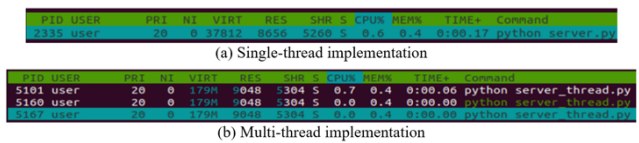

**FIGURE 21.** The processing and memory system resources obtained from htop -p <process_id> for the two distinct implementations of the cluster manager under comparison

### E. LIMITATIONS OF CURRENT WORK

Analyzing current work contribution, the runtime adaptation of resources at edge cloud systems seems a powerful strategy to mitigate against system disruptive events. Nevertheless, we can detect two important limitations on our research: i) proposal applicability to operation scenarios with high complexity, and ii) the centralized design of the cluster server. We aim in the future to investigate further enhancements on the current proposal to address the previous referred shortcomings.

### VI. DISCUSSION

We have subjected RECS to a carefully chosen set of tests, presented in Section V. From the results obtained, we can make the following observations.

We have shown that despite switches becoming disconnected from the SDN controllers, they can continue their normal switching operation because they operate as legacy Link Layer learning switches after the disconnection from the upper-level programmable switching logic entities.

We have identified the positive effects of a server farm but managed by SDN controllers in terms of the service quality provided to a high number of clients. We have also demonstrated that the servers inside the server farm can be activated in an elastic way according to the load variation. This could imply huge savings of energy by disconnecting unnecessary network equipment at off-peak hours. We have evidenced that Select groups at the data message forwarding level further enhance the quality of service provided by a server farm to its clients. The Select groups are also relevant to protect the performance of TCP traffic in cases there are competitive flows that monopolize the usage of available network resources.

The scheduling scheme that uses the Packet-In order number to orchestrate among the SDN controllers and decide which controller should process that Packet-In is fairer in terms of balancing the control load among the diverse SDN controllers than the alternative scheduling scheme; the other scheme uses the id of the switch where the Packet-In message was sent. However, the first scheduling scheme overloads the control channel more than the second one.

From our results, we can conclude that a multi-thread design for the cluster server is better than a single-thread alternative because the former implies less signaling traffic at the management channel. Nevertheless, the former consumes more processing resources than the latter. In addition, after an SDN controller failure, the measured system failover time is lower than 1ms.

Now we discuss the applicability of our proposal to real-life scenarios in two next steps. First, Internet Service Providers aiming to minimize the deployment and operational costs of their network infrastructure can benefit from using energy-aware solutions based on our proposal, which adjusts running infrastructure resources to the network demand evolution. Second, we believe our results represent a step forward in managing emerging edge applications that can benefit from low latency and high throughput. The typical upcoming edge applications are as follows: digital-valuable cases such as 360º video with virtual/augmented reality; critical control platforms; AI-enriched data and knowledge discovery systems; real-time e-commerce product recommendation; location-based multimedia; and heterogeneous information sharing among self-driving vehicles. These edge applications could be managed by a serverless service management platform [13], which would support service deployment, service discovery, or service life cycle management among other possible relevant features.

### VII. CONCLUSION AND FUTURE WORK

This paper introduces the design, deployment, and testing of RECS, a cross-level serverless edge-programmable solution to accomplish the goals of: i) detecting and remediating disconnections between SDN controllers and switches; ii) easing the burden of a congested server suffering from an over-demand of simultaneous client requests for its service; iii) mitigating the negative performance effects caused by congested network links; and iv) addressing failures and orchestrating multiple SDN controllers at the programmable switching logic using a control cluster manager together with a novel management protocol.

For future work, we aim to explore the usage of learning techniques at the cluster manager to decide how the SDN controllers should be orchestrated based on previously retrieved statistical information from both the control channel of each SDN controller and the controlled system. Another important direction is to investigate a set of cluster managers for greater resilience in a federated domain environment.

## ACKNOWLEDGMENTS
Jose Moura acknowledges the support given by Instituto de Telecomunicações, Lisbon, Portugal.

David Hutchison gratefully acknowledges the support of the Next Generation Converged Digital Infrastructure (NG-CDI) Prosperity Partnership Project funded by UK's EPSRC (project number EP/R004935/1) and British Telecom plc.

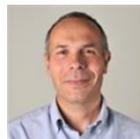

**Jose Moura** received the B.Sc. degree in Electronics and Telecommunications from Universidade de Aveiro (Portugal), M.Sc. in Computer Networks from Faculdade de Engenharia da Universidade do Porto (Portugal), and Ph.D. in Computer Science from Lancaster University (United Kingdom). From 1989 to 2000, he worked as Engineer at EFACEC Sistemas Electronica (Portugal) in Supervisory Control And Data Acquisition (SCADA) systems. From 2000 to 2001, he worked as a researcher at INESC-Porto (Portugal). Since 2001, he has been teaching in computer networks at ISCTE-Instituto Universitario Lisboa (Portugal), and he has been a researcher with Instituto de Telecomunicacoes (Portugal). He is an active reviewer for several Quartile 1 Journals covering his current research interests such as Network Management, Edge Computing, Optimization, Virtualization, Software-Defined Networking, and Resilience on Networked Systems.

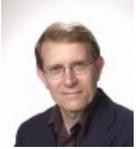
**David Hutchison** is Distinguished Professor of Computing at Lancaster University and founding Director of InfoLab21. He has served on the TPC of top conferences such as ACM SIGCOMM, IEEE Infocom, and served on editorial boards of the Springer Lecture Notes in Computer Science, Computer Networks Journal and IEEE TNSM, as well being editor of the Wiley book series in Computer Networks and Distributed Systems. He has helped build a strong research group in computer networks, which is well known internationally for contributions in a range of areas including Quality of Service architecture and mechanisms, multimedia caching and filtering, multicast engineering, active and programmable networking, content distribution networks, mobile IPv6 systems and applications, communications infrastructures for Grid based systems, testbed activities, and Internet Science. He now focuses largely on resilient and secure networking, with interests in Future Internet and the protection of critical infrastructures including industrial control systems.